\documentstyle[12pt,psfig]{article}
\newcommand{\beq}[1]{\begin{equation}\label{#1}}
\newcommand{\eeq}{\end{equation}}
\newcommand{\beqar}[1]{\begin{eqnarray}\label{#1}}
\newcommand{\eeqar}{\end{eqnarray}}
\newcommand{\lash}[1]{\not\! #1 \,}
\begin{document}
\vspace*{-2cm}
\hfill UFTP preprint 366/1994
\vspace{4cm}
\begin{center}
{\bf \large  QCD Sum Rule Calculation of Twist-3 Contributions to
Polarized Nucleon Structure Functions}
\vspace{1cm}

E.~Stein$^1$, P.~G\'ornicki$^2$, L.~Mankiewicz$^{3}$, A.~Sch\"afer$^1$
and W.~Greiner$^1$
\vspace{1cm}

$^1$Institut f\"ur Theoretische Physik, J.~W.~Goethe
Universit\"at Frankfurt,
\\
Postfach~11~19~32, W-60054~Frankfurt am Main, Germany
\\
$^2$Institute of Physics, Polish Academy of Sciences,
Al. Lotnikow 32/46, PL-02-668 Warsaw, Poland
\\
$^3$N. Copernicus Astronomical Center, Bartycka 18, PL--00--716 Warsaw,
Poland
\end{center}

\vspace{2cm}
\noindent{\bf Abstract:}

Using the framework of QCD sum rules we predict the twist-3
contribution to the second moment of the polarized nucleon structure
function $g_2(x)$. As the relevant local operator depends explicitely
on the gluon field, we employ a recently studied interpolating nucleon
current which contains three quark field and one
gluon field operator. Despite
the fact that our calculation is based on the analysis of a completely
different correlation function, our estimates are consitent with those
of Balitsky, Braun and Kolesnichenko who used a three-quark
current.
\eject
\newpage

Considerable attention has been paid to measurements of the polarized
structure function $g_1$ in deep inelastic scattering.  The unexpected
EMC \cite{EMC} results for the first moment of $g_1^p$, the structure
function of the proton, invoked tremendous effort to gain more
knowledge about polarized scattering.
As a result of these intense discussions it became clear that
polarization phenomena might offer much better opportunities to
test QCD than unpolarized experiments.
Meanwhile SMC \cite{SMC} and SLAC \cite{SLAC} experiments have
provided date on  $g_1^n$ and much improved ones on
$g_1^p$. Mainly due to this experiments the general interest shifted to
related questions most notably to the problem of $Q^2$ dependence.
It turned out that an
estimate of the magnitude of higher twist
contributions is urgently needed in order to compare the experiments
of SLAC, SMC and EMC
which cover different $Q^2$ regions \cite{Kar}. This is especially
important for the SLAC data which correspond to $Q^2$ around 2 ${\rm
GeV^2}$.
Besides being of great importance for the interpretation of the experiments
it became also clear that the leading higher matrix elements test fundamental
properties of the nucleon. Most notably $d^{(2)}$ on which we will focus in
this contribution is a measure for the mean colour-magnetic field in the
direction
of a transverse nucleon spin.
(Due to colour symmetry the expectation value $\langle \vec{B}^a\rangle = 0$,
but the expectation value of the colour singlet
$\langle \bar q B_1 q \rangle =  \langle \bar q \gamma_0 \tilde{G}_{01}
q\rangle
\sim d^{(2)}$ can be nonzero.)

The first moment of $g_1(x,Q^2)$ at fixed $Q^2$ is given by \cite{bruno}
\beq{eins}
\int_0^1 dx \; g_1 (x,Q^2)
=
\frac{1}{2} a^{(0)} + \frac{m_N^2}{9 Q^2}
\left( a^{(2)} + 4 d^{(2)} + 4 f^{(2)} \right) + O ( \frac{m_N^4}{Q^4} )
\; .
\eeq
The above formula does not include perturbative corrections (QCD).
Higher order corrections  were recently calculated in the
leading twist approximation up to
$O(\alpha_S^3)$ for non-singlet quantities, the Bjorken sum
rule and -- up
to $O(\alpha_S^2)$ -- for the Ellis-Jaffe sum rule \cite{larin}.
The Ellis-Jaffe sum rule has a flavour singlet contribution.

The reduced matrix elements for the twist-3 and twist-4
components of Eq. (\ref{eins}) are
defined by:
\beqar{zwei}
&&
\frac{1}{6} \langle pS| \bar{q} (0) \left[ \gamma^\alpha g
\tilde{G}^{\beta \sigma } + \gamma^\beta g \tilde{G}^{\alpha \sigma }
\right] q(0) |pS \rangle - {\rm traces } =
\nonumber  \\
&&=
2 d^{(2)} \left[ \frac{1}{6}
\left(
2 p^\alpha p^\beta S^\sigma + 2 p^\beta p^\alpha S^\sigma
- p^\beta p^\sigma S^\alpha - p^\alpha p^\sigma S^\beta
- p^\sigma p^\alpha S^\beta - p^\sigma p^\beta S^\alpha
\right)  \right.
\nonumber \\
&&
\enspace - {\rm traces } ]   \; ,
\eeqar
and
\beq{zwei1}
\langle pS| \bar{q}(0)
g \tilde{G}_{\alpha \beta} \gamma^\beta
q (0) |pS \rangle =
2 m_N^2 f^{(2)} S_\alpha \;
\eeq
respectively, where $|pS \rangle$ represents the nucleon state of
momentum $p$ and spin $S$, $S^2=-m_N^2$.

The $d^{(2)}$ matrix element does not only lead to a contribution
to the Ellis-Jaffe and Bjorken sum rules but is also in principle
measureable as the twist-3 contribution to the second moment of $g_2$
\cite{Jaffe1}:
\beq{g2moment}
\int_0^1 dx x^2 g_2(x,Q^2) = - \frac{2}{3}
\int_0^1 dx x^2 g_1(x,Q^2) + \frac{1}{3} d^{(2)} .
\eeq

Although both twist-3 and twist-4 matrix elements are interesting, in
the present paper we concentrate on the calculation of the twist-3
matrix element $d^{(2)}$ using the QCD sum rules. Some time ago
$d^{(2)}$ and $f^{(2)}$ were estimated with this technique by
Balitsky, Braun and Kolesnichenko \footnote{In their notation $d^{(2)}
= <<V>> / 4 $ and $f^{(2)} = -<<U>>/m_N^2$.} (BBK) \cite{BBK}.
They got rather suprising results, namely a strong asymmetry
between $d^{(2)}_{\rm BBK}({\rm proton})$ and
$d^{(2)}_{\rm BBK}({\rm neutron})$ and a $d^{(2)}_{\rm BBK}({\rm proton})$
which has opposite sign to that suggested by bag model calculations.
\cite{jiunrau}. It has been argued \cite{bruno2} that these qualitative
properties
should be observable in other spin-dependent effects, like single
spin asymmetries, such that their unambigious determination is of
great importance.
Any calculation of a non-perturbative quantity like $d^{(2)}$ has to
be based on a trustworthy well definded method which is insensitive to
technical
details and physicaly irrelevant assumptions.
In our opinion the only such available techniques are QCD sum rules
and lattice gauge calculations. The latter are tried but are still in
prelimenary
state. \cite{horsley}. Thus the reliability of the BBK result can only be
checked by consistency tests, namley by comparing
the results of calculations with substantially different interpolating
currents for which the systematic errors are uncorrelated.
The spread of obtained results gives a good estimate for the size of
the systematic errors involved.
There is actually reason for concern.
The chiral structure of the three-quark current used by BBK is such, that
graphs of a certain topology ``accidentally'' vanish because a trace of
an odd number of gamma-matrices has to be taken.
Unfortunately their sum rule is primarily determined by just a single graph
and
for this graph the up-quark contribution vanishes
due to the special structure of the used three-quark current.
Therefore the question arises if the obtained result reflects only the
the symmetry of
the choosen current or if there is some deeper physical reason
behind this.
In this situation independent
tests with different (more complicated) interpolating nucleon fields
are obviously necessary.

Unluckily the QCD sum rules technique used to estimate the matrix
elements present in
Eqs. (\ref{zwei}) and (\ref{zwei1}) leads to very complex computations.  As
the
consequence the fundamental results of BBK have not yet been checked.
Given their importance for the phenomenology we have decided to meet
the challenge and to perform an independent calculation with a
different interpolating current for the nucleon.

The starting point for calculating the reduced matrix elements of an
operator $O_\Gamma$ is the three-point correlation function
\beq{dreipunkt}
\Pi_\Gamma(p) = i^2 \int d^4x e^{ipx} \int d^4y
\langle 0|T\left\{ \eta(x) O_\Gamma(y) \overline{\eta}(0)\right\}| 0 \rangle
\eeq
where $\eta(x)$, $\langle 0| \eta(0)|pS\rangle = \lambda u(p,S)$ is an
interpolating current for the nucleon. There exists a certain
freedom of
choice for this current. While all reasonable choices should give
similar results the optimal choice must be tailored to the
problem considered. For a long time the standard
choice for sum rule calculations of nucleon properties has been the
three-quark current introduced by Ioffe \cite{Ioffe}
\beq{ioffe}
\eta_I(x) = \left[u^a(x) C \gamma_\mu u^b(x)\right]
                  \gamma_5 \gamma^\mu d^c(x) \varepsilon^{abc}.
\eeq
This current is used by BBK in their calculation.

As it can be seen from (\ref{zwei}) and (\ref{zwei1}), the operators
defining $d^{(2)}$ and $f^{(2)}$ explicitely contain the gluonic field
operator.  This gluonic field has to be matched by another gluonic
field operator. Because the three-quark current $\eta_I(x)$ does not
contain the gluonic operator explicitely the desired contribution must
be generated through the perturbative emission of an additional
gluon. The amplitude for this process is proportional to the strong
coupling constant $g$.

Alternatively, one may consider an interpolating field which in
addition to three quark fields contains the gluon field
explicitely. In that case gluon emission from the three-quark
configuration has a non-perturbative character.

A possible construction of an interpolating current with such
properties was discussed in \cite{BGMS}. It was noticed that all the
nice features of the current in (\ref{ioffe}) are preserved by making
it {\em nonlocal}, e.g. by shifting the $d$-quark to the point
$x+\epsilon$ (and adding the required gauge factor to preserve gauge
invariance). Expanding two times in $\epsilon $ and averaging over the
directions in Euclidean space $\epsilon_\mu \epsilon_\nu \rightarrow
\frac{1}{4} \delta_{\mu\nu} \epsilon^2 $ one arrives at
\beq{currG}
\eta^\prime_G(x)=
\varepsilon^{abc} (u^a(x)C\gamma_\mu u^b(x)) \gamma_5 \gamma_\mu
\sigma_{\alpha\beta}\left[ {\rm g} G^{\alpha\beta}(x) d(x)
\right ]^c \, .
\eeq
Finally, projecting out the isospin $\frac{1}{2}$ component leads to
the following proton interpolating current:
\beq{qqqG}
\eta_G(x) =
 \frac{2}{3} \left(\eta^{\rm old}_G(x) - \eta^{\rm ex}_G(x)\right) \; ,
\eeq
where
\beq{qqqGold}
\eta^{\rm old}_G(x) =\varepsilon^{abc} \left(u^a(x) C \gamma_\mu
u^b(x)\right)
\gamma_5 \gamma^\mu \sigma_{\alpha \beta} \left[G^{\alpha \beta}(x) d(x)
\right]^c \; ,
\eeq
and
\beq{qqqGex}
\eta^{\rm ex}_G(x) = \varepsilon^{abc} \left(u^a(x) C \gamma_\mu
d^b(x)\right)
\gamma_5 \gamma^\mu \sigma_{\alpha \beta} \left[G^{\alpha \beta}(x) u(x)
\right]^c \; .
\eeq

We would like to stress here that the technique of QCD sum rules does not
require the use of ``the best'' current from all the possible ones, it is
only necessary that the current is not too bad in order that the contribution
of interest is not suppressed by some special reason.

In ref \cite{BGMS} the current (\ref{qqqG}) was analysed in details and
the corresponding overlap $\lambda_G$ defined through
\beq{lambdaG}
\langle 0| \eta(0)|pS\rangle =
m_N^2 \lambda_G u(p,S) \; ,
\eeq
was calculated. Subsequently, it was used to
predict the nucleon momentum fraction carried by gluons to be $\sim
0.4$ at $\mu^2 \sim$ 1 GeV$^2$, and the gluon form factor of the
proton at $Q^2 \sim $ 2 -- 3 GeV$^2$. Both predictions agree very well
with present day data and its generally accepted interpretation.
This convinced us that the current
$\eta_G(x)$ is an adequate choice to calculate properties of the
nucleon.

 From the purely technical point of view the current $\eta_G(x)$ differs
in one important respect from the Ioffe current $\eta_I(x)$.
As mentioned before, the
chiral structure of $\eta_I(x)$ is such that graphs of certain
topologies vanish because a trace of an odd numbers
of gamma matrices has to be taken. The same happens obviously also for
$\eta^{\rm old}(x)$ but from correlators involving $\eta^{\rm ex}(x)$
we obtain non-vanishing contributions form the graphs of the same
topology. The calculation is therefore quite different from that
described in \cite{BBK}.

It is not yet possible to calculate the three-point correlator defined
by Eq. (\ref{dreipunkt}) directly. However, the usual philosophy of
the QCD sum rules is to relate it to some other correlators using
Wilson's operator product expansion (OPE). As it has been noticed in
\cite{Bal1} for the case of a three-point correlation function the
expansion is more complicated than in the case of two point functions
and has the following structure:
\beq{ope}
\Pi_\Gamma(p) =   \sum_n c_{\Gamma,n}^{\rm L}(p) <O^{\rm L}_n>
                  + \sum_n c^{\rm BL}_n(p) \Pi_{\Gamma,n}^{\rm BL}(0)
\; .
\eeq
As usual the main objective is to separate different scales:
vacuum expectation values of local operators $\langle O^L\rangle$
and correlators $\Pi^{BL}$ describe long distance effects while
the coefficients $c^L$ and $c^{BL}$ receive contribution only from highly
virtual quark and gluon fields which propagate for small distances.
The first part of the expansion -- so called local power
corrections (LPC) -- is well known from two point QCD sum rules.
The second part -- bilocal power corrections (BPC) -- has been
introduced in \cite{Bal1} to describe the large $y$ region of the integrand
in Eq. (\ref{dreipunkt}). This region cannot be properly described
by local power corrections alone. Yet, the distinction between large
and small distances is somewhat arbitrary and the corresponding
scale is implicitly included in the definitions of local and
bilocal power corrections. Changing the scale we can shift
part of the BPCs to LPCs and vice
versa. The optimal scale is always dictated by the nature of
the physical phenomena one wants to investigate.

The (OPE) of the correlation function $\Pi_\Gamma$ is given by the sum
of both LPC's and BPC's, as in (\ref{ope}). We note that in general
only this sum has a physical meaning and is independent of
the regularisation scheme.

The bilocal power corrections are determined by the long-distance,
non-perturbative contribution to correlation functions at zero
momentum \cite{Bal2}:
\beq{twopoint} \Pi_{\Gamma,n}^{\rm BL}(q) = i
\int d^4y e^{iqy} \langle 0|T\left\{O_\Gamma(y)
\tilde{O}_n(0)\right\}|0\rangle
\eeq with $q=0$. The Operators $\tilde{O}_n(0)$ are defined by the
following series
\beq{ope1}
T(\eta_I(x){\bar \eta}_G(0)) = \sum_n C_n^{BL}(x^2) {\tilde O}_n(0) ,
\eeq
where the above expansion goes over a series of local,
gauge-invariant operators of dimension $n$ ${\tilde O}_n(0)$. When this
expansion is inserted back in (\ref{dreipunkt}) it gives the second
contribution
to Eq.(\ref{ope}). We have performed a detailed analysis of the
BPC's to the sum rule for the twist-3 matrix element $d^{(2)}$ and have found
that numerically important correlators do not contribute due to symmetry.
In this respect the situation is entirely
the same as in the case of the calculation discussed in
\cite{BBK}.

The QCD sum rules for the twist-3 part of $g_1$ can be represented as
\beqar{twist3}
&&
\frac{1}{4} {\rm Tr}\left[
\lash{S} \gamma_5 i^2 \int d^4x e^{ipx} \int d^4y
\langle 0|T \left\{
\eta_I(x) \bar{q}(y) g\tilde{G}_{\sigma\{\mu}(y) \gamma_{\nu\}} q(y)
\eta_G(0)
\right\}|0\rangle\right] \nonumber \\
&&= -2 \frac{d^{(2)} \lambda_I \lambda_G
S_{[\sigma} p_{\{\mu]} p_{\nu\}} m_N^4}{(m_N^2 - p^2)^2}
+ \ldots
\eeqar
$[\ldots]$ means antisymmetrization of the indices, while $\{\ldots\}$
denotes symmetrization resp (see Eq. (\ref{zwei})).
Here, we have singled out the double-pole due to the nucleon contribution in
the intermediate state; other terms denoted by ellipses at the
phenomenological
right hand side are less singular in $m_N^2 - p^2$ and do not carry any
useful
information --
later on they will be suppressed by suitable treatment.
The couplings of the Ioffe current respectively our
quark - gluon current to the the nucleon are denoted by $\lambda_I$ and
$\lambda_G$
while $m_N$ denotes the nucleon mass.

The left hand side has been expanded and all LPC contributions
including operators up to dimension 10 have been calculated. It has
required calculation of more than 100 Feynman diagrams.
Traces of gamma matrices and other manipulations
were done using program TRACER \cite{tracer} written for
MATHEMATICA. Dimensional regularization has been used and the $\gamma_5$
matrix has been treated according to the 't Hooft-Veltman scheme
\cite{tHV}.

The result for the twist-3 contribution has the general form
\beqar{twist3ope}
&& {\rm Tr}\left[
\lash{S} \gamma_5 \Pi_{\sigma\mu\nu}(p)\right]
= \nonumber \\
&&
\left[ A \frac{\alpha_S}{\pi^5} (-p^2)^3 \log(-p^2/\mu^2)
+ B \frac{\alpha_S}{\pi} <\bar{q} q>^2 \log(-p^2/\mu^2)
\right . \nonumber \\
&&
+ C \frac{1}{\pi^4} <fg^3GGG> \log(-p^2/\mu^2)
\nonumber \\
&&
+ \left(D + F \log(-p^2/\mu^2)\right)
\frac{\alpha_S}{\pi} \frac{m_0^2<\bar{q} q>^2}{-p^2}
\nonumber \\
&& \left.
+ G \frac{<\bar{q} q>^2 <g^2G^2>}{(-p^2)^2}
+ H \frac{<\bar{q} q>^2 m_0^4}{(-p^2)^2}
\right] S_{[\sigma} p_{\{\mu]} p_{\nu\}}
\eeqar
where the numerical coefficients are given in Table \ref{tabelle}.
Note that the coefficient in front of the gluon condensate $\langle
GG \rangle$ vanishes.
\begin{table}
\begin{center}
\begin{tabular}{||l|l|l|l|l|l|l|l||}
\hline \hline
  &  $A$  &  $B$ & $C$ & $D$ & $F$ & $G$ & $H$ \\ \hline
d & $ 1/( 3^4 2^7) $ & 8/27  & $1/(3^2 2^6)$ & 85/162  & -95/648  & 10/81 &
2/9  \\ \hline
u & $ 1/( 3^4 2^6) $ & 4/9   & $1/(3^2 2^5)$ & -847/2592 & -13/324  &  2/81 &
-1/27 \\ \hline
\hline
\end{tabular}
\end{center}
\caption[]{Numerical coefficients corresponding to the sum rule eq.
\ref{twist3ope}.
The upper line gives the values for the twist-3 operator involving d quarks,
the lower line for the corresponding operator with u-quarks}
\label{tabelle}
\end{table}

There are two types of unwanted terms at the phenomenological
sides of the sum rules. The first type are the terms containing
a $(m_N^2 - p^2)^{-1}$ singularity.
The second are pure continuum contributions which correspond to
contributions from various
higher resonances etc.
To deal with these terms we first multiply both sides of
the sum rules by $(m_N^2-p^2)$ and then perform the Borel
transform and a suitable continuum subtraction.
The multiplication has a twofold effect: (a) it lowers the
singularity of the first term and (b) removes the $m_N^2 - p^2$
singularities from all other terms. In that way the only low
lying singularity is that of the first term. The
singularities at higher values of $-p^2 $ are suppressed by the
Borel transformation.
The transformed sum rule for twist-3  contribution
looks as follows:
\beqar{sumtwist3}
&&-16 d^{(2)} \lambda_I \lambda_G m_N^4 e^{-m_N^2/M^2}
\nonumber \\
&&= 3 ! M^{12} \tilde A (\frac{m_N^2}{M^2} E_4 - 4 E_5) +
        M^{6} (\tilde B + \tilde C) (E_2 - \frac{m_N^2}{M^2} E_1) +
        M^2 m_N^2 \tilde D +
\nonumber \\
&&      M^4 \tilde F (\frac{m_N^2}{M^2} \log(M^2/\mu^2) - E_1) +
        M^2(\tilde G + \tilde H)(\frac{m_N^2}{M^2} + 1)
\eeqar
where
\beq{eformel}
E_{n} = 1 - e^{\left(-s_0/ M^2 \right)} \sum_{k = 0}^{n-1}
\frac{1}{k!}\left(s_0 / M^2\right)^k
\eeq
Here we absorbed the condensates, $\alpha_S$ and $\pi$ factors in the
definition of the coefficients $\tilde A, \tilde B , \ldots$ which then are
given by
$\tilde A = \alpha_S A / \pi^5$,
$\tilde B = \alpha_S <\bar{q} q>^2 B / \pi$,
$\tilde C = <fg^3GGG> C / \pi^4$,
$\tilde D = \alpha_S m_0^2 <\bar{q} q>^2 D / \pi$,
$\tilde F = \alpha_S m_0^2 <\bar{q} q>^2 F / \pi$,
$\tilde G = <g^2 G^2> <\bar{q} q>^2 G $,
$\tilde H = m_0^4 <\bar{q} q>^2 H $.
\\
One point requires explanation here. The term associated with
the dimension-8 condensate in the sum rule for the matrix element of
the twist-3 operator contains arbitrary scale $\mu$ even after Borel
transformation. This dependence may be traced back to the occurence of
terms proportional to $\log(-p^2/\mu^2)/p^2$ in the original sum
rule. They arise because of the mixing of the three-point correlation
function (\ref{dreipunkt}) with two-point correlation functions of the
nucleon currents $\eta_I(x)$ and ${\bar \eta}_G(0)$ with operators
produced by the contraction of $O_\Gamma(y)$ with ${\bar \eta}_G(0)$
and $\eta_I(x)$, respectively \cite{BK}.  The natural choice of the
scale is $\mu\approx$ 1 GeV which corresponds to the typical values of
momenta in the intermediate region where the sum rule equality is
expected to hold.  Fortunately the numerical coefficient associated
with this term is small.

The condensates have been normalized at 1 GeV. Their values:
$<\bar{q}q> = (-0.257 {\rm GeV})^3$,
$<\alpha_s/\pi GG> = 0.012 {\rm GeV}^4$,
$m_0^2 = <\bar{q} g\sigma Gq>/ <\bar q q> = 0.65 {\rm GeV^2}$
and $<fg^3GGG> = 0.046 {\rm GeV^6}$
correspond to the standard ITEP values rescaled to the normalization point
$\mu^2 \sim m_N^2 \sim 1 \;{\rm GeV^2}$.
The strong coupling at $1 \;{\rm GeV}$ is taken to be
$\alpha_S = 0.37$ ($\Lambda = 150)$ MeV.
The continuum
threshold is taken at $s_0=(1.5{\rm GeV})^2$ roughly corresponding to
the Roper resonance position.

The coupling constants $\lambda_I$ and $\lambda_G$ may be
calculated from the two point sum rule derived in \cite{BGMS}
\beqar{normierung}
&&2 (2 \pi)^4 m_N^2 \lambda_I\lambda_G  e^{-m_N^2/M^2}
= \frac{6}{5} \frac{\alpha_S}{\pi} M^8 E_4  \nonumber \\
&& + \frac{1}{2} <g^2GG> M^4E_2
   -\frac{4}{3}\frac{\alpha_s}{\pi}(2\pi)^4<\bar q q>^2 M^2E_1
   + \frac{2}{3}(2\pi)^4m_0^2<\bar q q>^2 \nonumber \\
\eeqar
\begin{figure}
\centerline{\psfig{figure=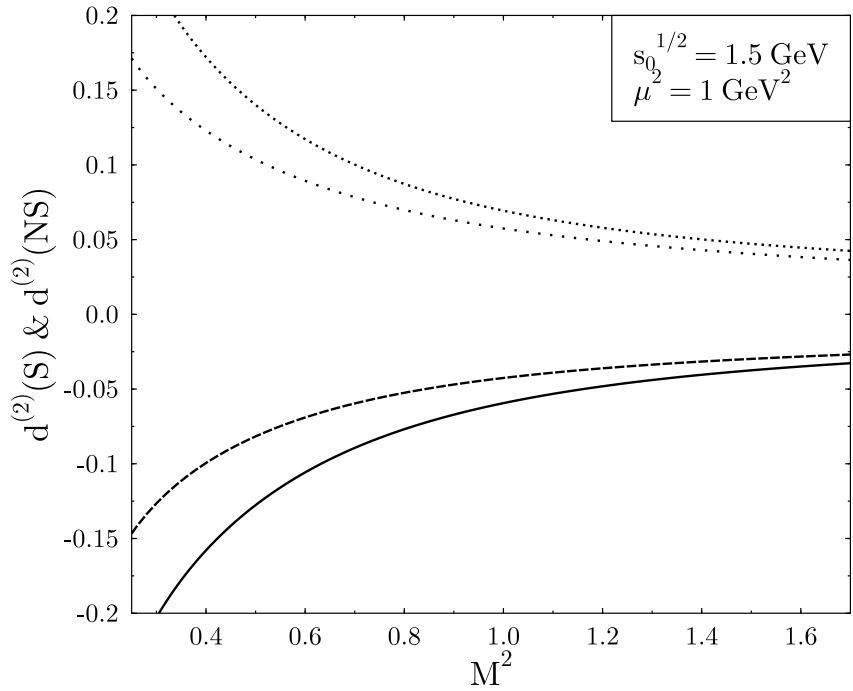,width=12cm}}
\caption[]{Stability plot of the sum rule eq. (\ref{sumtwist3}).
The full line corresponds to $d^{(2)}(S)$, the
dotted line to $d^{(2)}(NS)$. For comparison the results of the analysis
in \cite{BBK} are also shown. The dashed line corresponds to
$d^{(2)}_{BBK}(S)$
the space-dotted line to $d^{(2)}_{BBK}(NS)$.}
\label{fig1}
\end{figure}
\begin{figure}
\centerline{\psfig{figure=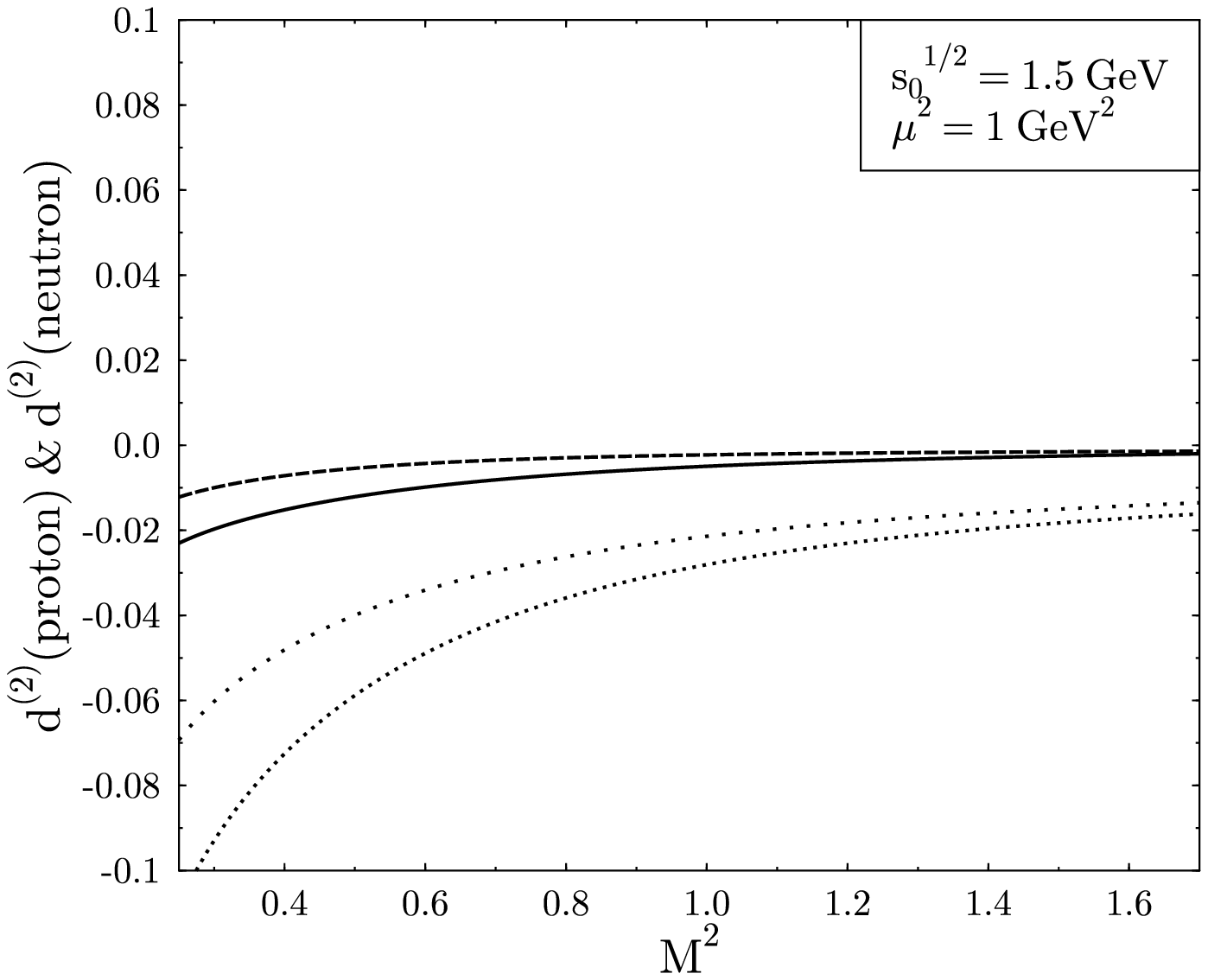,width=12cm}}
\caption[]{Dependence of $d^{(2)}(proton)$ and
$d^{(2)}(neutron)$ on the Borel parameter $M^2$.
The full line correponds to the $d^{(2)}_p$ the dotted line to
$d^{(2)}_n$. The dashed and space-dotted lines correspond to the
results for $d^{(2)}_{\rm BBK}(proton)$ and
$d^{(2)}_{\rm BBK}(neutron)$ obtained from the
sum rules given in \cite{BBK}.}
\label{fig2}
\end{figure}
However, it is more convenient (and usually more accurate) to divide simply
the sum rule (\ref{sumtwist3}) by the sum rule (\ref{normierung}). The
coupling constants at the phenomenological side cancel out. The
quotient sum rule for the matrix element of twist-3 operator has been
plotted in Fig.  \ref{fig1}. This figure shows the singlet (S) and
nonsinglet (NS) part of $d^{(2)}$.
$d^{(2)}(S) = d^{(2)}(u) + d^{(2)}(d)$,
$d^{(2)}(NS) = d^{(2)}(u) - d^{(2)}(d)$.
The corresponding matrix elements $d^{(2)}(proton)$ and $d^{(2)}(neutron)$
are shown in Fig. \ref{fig2}.
For comparison the corresponding sum rules obtained from the the analysis
in \cite{BBK} which employed Ioffe currents only are shown.
The corresponding coupling constant $\lambda_I^2$ is determined from
the additional two point sum rule
\beqar{ioffenorm}
&&2 (2 \pi)^4  \lambda_I^2  e^{-m_N^2/M^2}
=  M^6 E_3 \nonumber \\
&& + \frac{1}{4} <g^2GG> M^2E_1  + \frac{4}{3}(2\pi)^4<\bar q q>^2
\eeqar
which is just the standard sum rule considered by Ioffe.
Instead of using a fixed value for  $\lambda_I^2$
we divided the sum rule obtained in \cite{BBK} by the sum rule
(\ref{ioffenorm}) which improved the stability considerably.
The final values of the matrix elements  may then be estimated from the
figures
by taking $M^2\approx m_N^2$:
\beqar{numbers}
&&d^{(2)}(S) = - 0.068 \pm 0.03   \quad
d^{(2)}(NS) = 0.078 \pm 0.03
\nonumber \\
&&d^{(2)}(proton) = -0.006 \pm 0.003  \nonumber \\
&&d^{(2)}(neutron) =  -0.03 \pm  0.01
\eeqar
Errors are due to the $M^2$ dependence of the sum rule, see Figure
\ref{fig1} and \ref{fig2} resp.
The uncertainty introduced by the hypothesis of factorization of
higher-dimension condensates is discussed below.
These values are to be compared with those obtained from the sum rules
given in \cite{BBK}:
\beqar{numbersBBK}
&&d^{(2)}_{\rm BBK}(S) = - 0.05 \pm 0.03 \quad
d^{(2)}_{\rm BBK}(NS)  = 0.065 \pm 0.03
\nonumber \\
&& d^{(2)}_{\rm BBK}(proton) = - 0.003 \pm 0.003 \nonumber \\
&& d^{(2)}_{\rm BBK}(neutron)  = -0.025 \pm  0.01 \nonumber
\eeqar

In a recent paper \cite{Ross} Ross and Roberts have estimated
contributions to the sum rule (\ref{twist3}) arising from chiral
corrections to the phenomenological spectral density.
However it is not clear how to take into account the corresponding
chiral contributions on the theoretical side of (\ref{twist3}), see
e.g. \cite{griegel}.
When the necessary ``theoretical error'' is taken
into account, their result agrees
with that of \cite{BBK}.
As this error is unavoidably large also in the present
case we have not applied the Ross and Roberts analysis to our results.

As in \cite{BBK} and \cite{BK} our sum rule is dominated by the
contribution from the highest dimension operators considered i.e.,
those with coefficients determined to the leading accuracy by tree
diagrams. As discussed in \cite{BK} this could in principal
signal a
breakdown of the OPE for the correlator in question which may not
allow to make a reliable numerical estimate. In \cite{BK} this
potentially important problem was settled by an estimate of the
contribution from operators of yet higher dimension 10. It was found
that the contribution of operators of dimension 10 is likely much
smaller than those of dimension 8, which dominated the sum rule in
\cite{BK}.
This is a situation very typical for QCD sum rules.
Low-dimension operators are  usually  numerically unimportant
because each extra loop in the definition of the coefficient in front
of an operator of lower dimension brings in a small factor $\sim (2
\pi^2)^{-1}$. On the other hand the higher  the dimension the stronger is
the suppression due to the Borel transfromation.
The dominant graphs are therefore usually the lowest dimension tree graphs.
In the present case
these are the dimension 10 contributions.
We expect that as usually still higher dimension contributions starting with
dimenision 12 will be small due to the Borel
transformation.
Unfortunately a quantitative test of these arguments
is difficult as very little is known about VEV of operators of
dimension 12.

Note also that our calculation involves factorization of the
condensate of high dimension. In this respect it is also very similar
to the calculation presented in ref. \cite{BBK}. The factorization of
the higher dimensional condensates is a standard technique used in the
QCD sum rules. It is necessary to reduce the number of unknown
parameters. The physical assumption hidden behind this procedure is
the absence of certain higher order correlations in the QCD
vacuum. The validity of this assumption has been studied for several
simpler condensates \cite{ZH}. Yet, the knowledge of condensates of
dimension 10 is scarce because such condensates rarely appear in
calculations.
One can even turn this argument around and argue
that the good agreement with the previous
calculation \cite{BBK} supports the validity of factorization
hypothesis. The BBK result may be treated as a kind of cross
check because it relies on a different condensate for the dominant
contribution, namely $<\bar{q}q>^2 m_0^2$ instead of
$<\bar{q}q>^2 <G^2>$ and $<\bar{q}q>^2 m_0^4$.
The other test is the two point sum rule for the
nucleon mass with two $\eta_G$ currents \cite{BGMS}. This calculation
involves condensates of the dimension 10 and leads to reasonable
results.

The agreement between the two calculations is a highly non-trivial
check of the QCD sum rules method.  The main difference is that in
\cite{BBK} gluons were generated entirely by perturbation theory while
in our case they are already present in the interpolating current. As
a consequence our correlator (\ref{dreipunkt}) has milder singularities
than that analysed in \cite{BBK}. In principle this should be an
advantage of our approach but as the current $\eta_G(x)$ has two units
higher dimension
than that of the $\eta_I(x)$, the accuracy is
diminished by the need to consider higher dimensional condensates. Given
an approximate character of the method it is by no means obvious that both
approaches give the same results. It seems to us that our results not only
give a strong argument in favour of the numerical estimates presented in
(\ref{numbers}) and (\ref{numbersBBK}), but also give a strong support to
the credibility of QCD sum rules method.
However, neither type of sum rule can be expected to yield very accurate
results, because of the slow convergence of the operator product expansion.

We think that taking all the uncertainties into account the following
three qualitative statements can be made:
\begin{itemize}
\item $|d^{(2)}_p|$ and  $|d^{(2)}_n|$ are smaller than 0.03.
The corrections from the twist-3 operator to
the Bjorken sum rule and Ellis-Jaffe sum rule are therefore
of order $0.013 m_N^2/ Q^2$ which e.g. for
$\int g_1^p(x) dx$ amounts to a correction of $\sim 5 \%$ at
$Q^2 \sim 2 {\rm GeV^2}$. Note that for a definite answer also
$a^{(2)}$ and $f^{(2)}$ have to be taken into account.

\item  There is a strong isospin asymmetry between $d^{(2)}_p$ and
$d^{(2)}_n$:  $|d^{(2)}_p| > |d^{(2)}_n|$.
This isospin asymmetry should manifest itself in a number
of spin phenomena \cite{bruno2}.

\item the sign of $d^{(2)}_p$ is negative (in contradiction to the
bag model prediction \cite{jiunrau}). $d^{(2)}_n$ is negative.
\end{itemize}

{\it Acknowledgements.} We would like to thank Vladimir Braun
for many usefull discussions and continous encouragement.
This work has been supported by KBN grant 2~P302~143~06.
A.S. thanks DFG (G.Hess Programm) and MPI f\"ur Kernphysik in Heidelberg
for support.
\vfill
\eject

\end{document}